\documentclass[aps,prb,twocolumn,showpacs,superscriptaddress,amsmath,amssymb]{revtex4-1} 

\usepackage{amsmath,amssymb}
\usepackage{ulem}
\usepackage{color}
\usepackage[colorlinks,citecolor=blue,urlcolor=blue]{hyperref}
\usepackage[capitalise]{cleveref}
\usepackage{graphicx}
\def \beq {\begin{equation}}
\def \edq {\end{equation}}
\def \bes {\begin{subequations}}
\def \eds {\end{subequations}}
\def \beqn {\begin{equation*}}
\def \edqn {\end{equation*}}

\def \veps {\varepsilon}

\def \rmL {{\rm{L}}}
\def \rmR {{\rm{R}}}

\newcommand{\bea}{\begin{eqnarray}}
\newcommand{\eea}{\end{eqnarray}}
\newcommand{\be}{\begin{equation}}
\newcommand{\ee}{\end{equation}}

\newcommand{\rme}{\mathrm{e}}

\begin{document}
\title{Dynamical Coulomb blockade of thermal transport}
\author{Guillem Rossell\'o}
  \affiliation{Institut de F\'{i}sica Interdisciplinar i de Sistemes Complexos
  IFISC (CSIC-UIB), E-07122 Palma de Mallorca, Spain}
\author{Rosa L\'opez}
\affiliation{Institut de F\'{i}sica Interdisciplinar i de Sistemes Complexos
  IFISC (CSIC-UIB), E-07122 Palma de Mallorca, Spain}
\author{Rafael S\'anchez}
\affiliation{ Instituto Gregorio Mill\'an, Universidad Carlos III de Madrid, 28911 Legan\'es, Madrid, Spain}
\date{\today}

\begin{abstract}
 The role of energy exchange between a quantum system and its environment is investigated from the perspective of the Onsager conductance matrix. We consider the thermoelectric linear transport of an interacting quantum dot coupled to two terminals under the influence of an electrical potential and a thermal bias. We implement in our model the effect of coupling to electromagnetic environmental modes created by nearby electrons within the $P(E)$-theory of dynamical Coulomb blockade. Our findings relate the lack of some symmetries among the Onsager matrix coefficients with an enhancement of the efficiency at maximum power and the occurrence of the heat rectification phenomenon. 
\end{abstract}

%\pacs{
%  73.63.-b, % Condensed Matter: Electronic Structure, Electrical, Magnetic, ...
            % - Electronic structure and electrical properties of ...
            % - Electronic transport in nanoscale materials and structure
%  74.50.+r, % Condensed Matter: Electronic Structure, Electrical, Magnetic, ...
            % - Superconductivity
            % - Tunneling phenomena: point contacts, weak links, Josephson effects
%  72.15.Qm, % - Scattering mechanisms and Kondo effect
%  73.63.Kv  % Condensed Matter: Electronic Structure, Electrical, Magnetic, ...
            % - Electronic structure and electrical properties of ...
            % - Quantum dots
%}
\maketitle

\section{Introduction}
\label{sec:introduction}

The properties of electronic heat transport in nanostructures have recently attracted the attention of the scientific community for different reasons~\cite{Giazotto2006,Dubi2011,Benenti2016}. On one hand, the onset of quantum effects in the mesoscopic regime opens the way to the investigation of the impact of quantum mechanics on thermodynamics.~\cite{Pekola2015} In particular, heat engines based on purely quantum mechanical effects have been recently proposed.~\cite{Sanchez2015b,Hofer2015,Hofer2016,Marchegiani2016,Rossnagel2016,Karimi2016}
%The particular properties of these systems and their achieved tunability postulate them as efficient heat engines, including  refrigerators~\cite{Giazotto2006,Prance2009}, thermoelectric generation~\cite{Dubi2011,Sothmann2015}, or thermal rectifiers~\cite{Scheibner2008}. Different ways to improve the thermoelectric performance by  using designed configurations of quantum dots~\cite{Sanchez2011a}, nanowires, heterostructures... are subject of intense research. Additionally, nonlinear transport permits to conceive thermal devices that manipulate heat by defining the operation of thermal diodes~\cite{Ruokola2011} and transitors.
%
%Quantum thermodynamics fundamentals and its potential applications have attracted much the attention of the scientific community in the last years.~\cite{Pekola2015} 
Complementary to this, there has been a spectacular progress in the field of quantum thermoelectrics, both from the theoretical and experimental sides. Exciting proposals like nanoprobe thermometers~\cite{Giazotto2006,Meair2014}, energy harvesting devices~\cite{Jordan2013,Donsa2014,Sothmann2015,Roche2015,Hartmann2015}, refrigerators~\cite{Nahum1994,Leivo1996,Prance2009,Koski2015,Kafanov2009}, heat diodes~\cite{Ruokola2011}, rectifiers~\cite{Segal2008,Ruokola2009,Thierschmann2015a,Sierra2015}, transistors~\cite{Joulain2016,Sanchez2017}, multi-terminal heat engines~\cite{Entin-Wohlman2010,Sanchez2011a,Matthews2012,Sanchez2015a} among others have come up in the last years. 

In this respect, quantum dots~\cite{Staring1993,Dzurak1993,Dzurak1997, Humphrey2002,Scheibner2007,Scheibner2008,Svensson2012,
Svensson2013,Thierschmann2015,Thierschmann2016} have a prominent role for being good energy filters that improve the thermoelectric efficiency~\cite{Mahan1996,Esposito2009}. The presence of strong interactions introduces the Coulomb blockade regime where transport can be controlled at the level of single-electron tunneling events~\cite{Averin1986,beenakker1992}. Different functionalities such as heat engines~\cite{Sothmann2015}, pumps~\cite{Rey2007,Juergens2013,Roche2013} and diodes~\cite{Ruokola2011} can be defined that use these properties.
%\sout{Efficiencies have been also deeply investigated in quantum systems within the linear and nonlinear regime~\cite{Whitney2015, Whitney2014, Entin-Wohlman2014, Yamamoto2016} and in situations when time reversal symmetry is broken by the presence of magnetic fields~\cite{Benenti2011a,Sanchez2011,Jiang2012,Brandner2013}. }

%Heat engines based on purely quantum mechanical effects have been recently proposed.~\cite{Sanchez2015a,Sanchez2015b,Hofer2015,Hofer2016,Marchegiani2016} 
Currents are small in nanostructures, and are hence sensible to fluctuations.
The question arises of how the system behaviour is influenced by a noisy environment. On one hand, it leads to dephasing and decoherence which are detrimental to quantum coherent processes. This is however not necessarily negative~\cite{Sanchez2016,Schiegg2016}. On the other hand, they may lead to inelastic transitions which can contribute to the engine performance by injecting or releasing energy in the conductor~\cite{Arguello2015,Jiang2015,Yamamoto2016}. Indeed, non-local thermoelectric engines exist that use an environment as a heat source in an otherwise equilibrated conductor. The nature of the  environment can either be fermionic~\cite{Sanchez2011,Sothmann2012a} or bosonic~\cite{Entin-Wohlman2010,Sothmann2012}. It can also consist of transport fluctuations in a Coulomb coupled conductor\cite{Sanchez2010,Bischoff2015, Kaasbjerg2016,Keller2016} or be due to quantum fluctuations in an electromagnetic environment\cite{Ruokola2012,Henriet2015}. This last effect has been observed in the form of the dynamical Coulomb blockade of charge currents~\cite{Delsing1989,Geerligs1989,Devoret1990,Girvin1990,Grabert1991}.

The linear response of a two terminal nanodevice is defined on the grounds of the Onsager-Casimir relations~\cite{Casimir1945,Callen1998,Jacquod2012}. The Onsager coefficients $\mathcal{O}_{ij}=\partial\mathcal{I}_i/\partial\mathcal{A}_j$ relate the charge and heat fluxes $\mathcal{I}_i\equiv\{I_i^e, I_i^h \}$ in terminal $i$ to the thermodynamic affinities $\mathcal{A}_j$. These can be due to electric or thermal gradients. Shortly, the coefficients $\mathcal{O}_{ij}$ can be collected into the so-called Onsager conductance matrix which is a symmetric and positively semi-defined matrix.~\cite{Onsager1931,*Onsager1931a} Derived from the principle of microreversivility, Onsager reciprocity relations identify non diagonal coefficients $\mathcal{O}_{ij}$, e.g. Seebeck and Peltier responses. Notoriously, such relations are also satisfied for quantum systems independently on the presence of interactions. For quantum systems in which phase coherence is preserved, additional relations for the Onsager coefficients are obtained from the unitarity of the electron dynamics~\cite{Buttiker1988,Butcher1990} giving rise to  highly symmetric Onsager matrices.  However, environment-system energy exchange events prevent the dynamics to be unitary. In particular some relations among the thermoelectric coefficients are no longer satisfied.~\cite{Sanchez2011,Saito2011} The microscopic origin of such asymmetries has so far not been discussed.%The same happens for the $K$ matrix as shown below. 

\begin{figure}[b]
\begin{center}
  \includegraphics[width=\linewidth,clip]{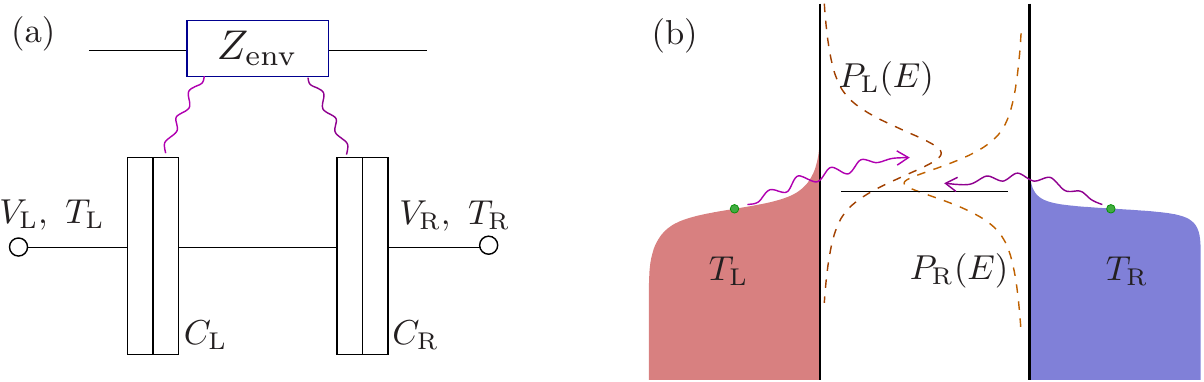}%}
\end{center}
\caption{\label{fig:scheme}(a) Schematic of a two terminal single-level quantum dot device in the presence of an electromagnetic environment described by an external impedance $Z_{\rm env}(\omega)$. Each terminal is electrically and thermally biased with $V_\rmL$, $T_\rmL=T_0 + \Delta T_\rmL$ (left contact) and $V_\rmR$, $T_\rmR=T_0+\Delta T_\rmR$ (right contact). The capacitance associated  to each junction, $C_i$, determines the dynamical coupling to the environment. (b) Inelastic tunneling into the quantum dot is then described by a distribution $P_j(E)$ which is different for each barrier.}
\end{figure}

In this work we explore this issue by using a microscopic model for the coupling of a mesoscopic system to a dynamical environment~\cite{Ingold1992}. We consider the simplest situation of a single-level quantum dot tunnel-coupled to two terminals, L and R. Particle tunneling between two quantum states is  accomplished by energy exchange with an electromagnetic environment that describes the vacuum fluctuations. The environment is modeled  by an external impedance, as sketched in Fig.~\ref{fig:scheme}. This results in photon-assisted tunneling events which on top of becoming inelastic, introduce left-right asymmetric rates. As we show below, the occurrence of Onsager matrix asymmetries is due to combination of these two effects. It leads to responses that do not only depend on the global temperature gradient $\Delta T=T_\rmL-T_\rmR$ but rather on how it is distributed with respect to some reference temperature $T_0$ in the two leads, $\Delta T_l=T_l-T_0$. It affects the thermoelectric response and most particularly, it introduces an apparent thermal rectification in the linear regime.

The remaining of the manuscript is organized as follows. In Sec.~\ref{sec:model} our model is described. Results for the single and double occupation regimes are presented in Secs.~\ref{sec:spinless} and \ref{sec:spinful}, respectively, with conclusions discussed in Sec.~\ref{sec:conclusions}. 

\section{Theoretical Model}
\label{sec:model}
We consider a two terminal interacting conductor as illustrated in \cref{fig:scheme}.  We use a spinful single  level quantum dot described by four states $|0\rangle $,  $|{\rm u}\rangle$, $|{\rm d}\rangle$ and $|2\rangle$. They correspond to an empty dot ($|0\rangle $), a singly occupied dot with either spin up ($|{\rm u}\rangle$) or down ($|{\rm d}\rangle$) polarization, and the doubly occupied dot state ($|2\rangle$). Our transport description is restricted to the sequential tunneling regime for which  $\Gamma \ll k_BT_0$  ($T_0$ is the temperature).  In this regime, transport events are predominantly of the first order in the tunneling coupling $\Gamma$. To properly account for Coulomb interactions we employ the electrostatic model schematically illustrated in \cref{fig:scheme}. In such model, the electrostatic charging energy is described with two capacitances $C_\rmL$ and $C_\rmR$.   The dynamics of the system is modeled by the time evolution of the occupation probabilities $p \equiv \{p_0,p_{\rm u},p_{\rm d},p_2\}$ described by the general master equation 
\begin{equation}
%\frac{dp}{dt} = \call_{\rm seq} p\,.
\frac{dp_i}{dt} = \sum_k\left(\Gamma_{i\leftarrow k}p_k-\Gamma_{k\leftarrow i}p_i\right).
\end{equation}
Applied to our formalism this equation would read e.g. for $p_0$:
\begin{equation}
%\frac{dp_0}{dt}=-(\Gamma_{\rm L,0}^+ + \Gamma_{\rm R,0}^+)p_0+(\Gamma_{\rm L,0}^- + \Gamma_{\rm R,0}^-) p_u +(\Gamma_{\rm L,0}^- + \Gamma_{\rm R,0}^-)p_d\,,
\frac{dp_0}{dt}=\sum_j\left[-\Gamma_{\rm j,0}^+p_0+\Gamma_{j,0}^- (p_{\rm u} +p_{\rm d})\right]\,,
\end{equation} 
 %that conforms the first row in $\call_{\rm seq}$.  Here $\Gamma_{\sigma,s}^{\pm} = \sum_{j} \Gamma_{j\sigma,s}^{\pm}$ with $j=\rmL,\rmR$ are the transition rates for the tunneling events involving the $s$ dot charge state  and the lead $j$. 
whose transition rates are of the form $\Gamma_{j,s}^{\pm}$ for electrons tunneling {\it in} (-) or {\it out} (+) of the dot through contact $j$ are given below. We do not consider a magnetic field, so they do not depend on spin.
%Individual transition rates like $\Gamma_{j\sigma,n}^{\pm}$ describe the $\sigma$-electron hop {\it in} (-) or {\it out} (+)  of electrons  through contact $j$. In the absence of environmental effects such \textit{in} and {\it out} transitions rates are simply given by 
% \begin{equation}
%  \Gamma_{j\sigma,s}^{\pm}=\Gamma_j f_{js}^{\pm}(\varepsilon_{d\sigma}-\mu_{\sigma,s})\,,
% \end{equation}
%where $f_{j}^+(E)=1/[1+e^{\beta_j(E-eV_j)}]$ is the Fermi function with $\beta_j=1/k_B T_j$,  $f_{j}^-=1-f_{j}^+$, and $\mu_{\sigma,s}$ is the dot electrochemical potential when the dot is empty $s=0$ or singly occupied $s=1$.  Our electrostatic model (see \cref{fig:scheme}) yields  
They depend on the electrochemical potential $\mu_{s}$ when the dot is empty $s=0$ or singly occupied $s=1$. A simple electrostatic electrostatic model~\cite{Beenakker1991} yields  
\begin{equation}
\mu_{s} = \epsilon_{\rm d} + \frac{e^2 (1+2 s)}{2C} +e(\kappa_\rmL V_\rmL + \kappa_\rmR V_\rmR)\,,
\end{equation}
where $\epsilon_{\rm d}$ is the bare energy level of the quantum dot, $C=C_\rmL+C_\rmR$ is its total capacitance, and $\kappa_j=1-C_j/C$, with $j=\rmL,\rmR$, see \cref{fig:scheme}. 

\subsection{Tunneling rates}

Tunneling events are frequently affected by fluctuations of the electromagnetic environment~\cite{Grabert1991}. To fully account for such quantum fluctuations we adopt the $P(E)$ theory~\cite{Grabert1991, Ingold1992} of dynamical Coulomb blockade, recently revisited to consider heat fluxes~\cite{Ruokola2012}. The spirit of the $P(E)$ theory relies on the fact that individual tunneling events involve energy exchange processes. The Dirac-delta accounting for energy conservation in the (Fermi golden rule) tunneling rates is relaxed into a brodened distribution $P(E)$.  More  specifically for a double junction it reads
\begin{equation}\label{pofe2}
P_j(E)=\frac{1}{2\pi \hbar}\int dt  \exp{\left(\kappa_i^2 J(t)+\frac{i}{\hbar} Et\right)}\,.
\end{equation}
where the function
\begin{equation}
J(t)=\frac{2h}{e^2}\int^{\infty}_0 \frac{d\omega}{\omega} Re[\tilde{Z}(\omega)]c(\omega,T_0)\,,
\end{equation} 
contains all the information of the environment fluctuations, with\cite{Ingold1992}:
\begin{equation}
c(\omega, T_0)=\coth\left(\frac{\hbar \omega}{2k_{\rm B}T_0}\right)[\cos(\omega t-1)-i\sin\omega t].
\end{equation}

If we consider a pure resistive or ohmic environment, i.e. we have
 \begin{equation}
 Z_{env}(\omega)=R\gg R_q=h/2e^2.
 \end{equation} 
This situation corresponds to the case where the electron tunnel may easily excite many electromagnetic modes.  Thus, the total impedance seen by the external circuit is 
\begin{equation}
\tilde{Z}(\omega)=[i\omega C_{\rm eff}+Z_{env}(\omega)]^{-1},
\end{equation}
with the effective capacitance of the quantum dot $C_{\rm eff}^{-1}=C_\rmL^{-1}+C_\rmR^{-1}$. Under these considerations the high impedance limit leads to a Gaussian distribution for $P_j(E)$:
\be
\label{PjofE}
P_j(E)=\frac{1}{\sqrt{4\pi\kappa_j^2E_C k_B T_{0}}}e^{-\frac{\left(E-\kappa_j^2E_C\right)^2}{4\kappa_j^2E_C k_{\rm B} T_{0}}},
\ee
Here $E_C=q^2\kappa_\rmL\kappa_\rmR/2C$. Remarkably, asymmetries in the system capacitances translate in the $P_j(E)$ functions having different mean, $\kappa_j^2E_C$, and variance, $2\kappa_j^2E_C k_B T_{0}$. They modify the transition rates expressions according to
 \begin{equation}\label{gammarates}
\Gamma^{\pm}_{j,s}=\Gamma_{j} \int dE f^{\pm}(E-eV_j,T_j) P_j(E-\mu_{s}).
\end{equation}
where $f^+(E,T)=1/[1+e^{E/(k_{\rm B}T)}]$ is the Fermi function, and $f^-=1-f^+$. The tunneling rates can be left-right asymmetric for having barriers with different transparencies, $\Gamma_\rmL\neq\Gamma_\rmR$. We emphasize that, very differently, having $P_\rmL(E)\neq P_\rmR(E)$ introduces an implicit energy dependence. As we discuss below, the impact in the system response shows up in the thermal transport coefficients.

Finally, the charge and heat currents through contact $j$ are calculated through
\begin{align}\label{currents}
%&I_{j}^{\rm e} = e \left( \Gamma_{j\sigma,0}^+ p_0 -\Gamma_{j\sigma,0}^- p_{\sigma} + \Gamma_{j\sigma,1}^+ p_{\bar\sigma} -\Gamma_{j\sigma,1}^- p_{2} \right)\,,\\ \nonumber
%&I_{j }^{\rm h} =\gamma_{j\sigma,0}^+ p_0 -\gamma_{j\sigma,0}^- p_{\sigma} + \gamma_{j\sigma,1}^+ p_{\bar\sigma} -\gamma_{j\sigma,1}^- p_{2}.
&I_{j}^e = e \left[\sum_{\sigma={\rm u, d}}\left(\Gamma_{j,0}^+ -\Gamma_{j,1}^-\right)p_\sigma -\Gamma_{j,0}^- p_{0} + \Gamma_{j,1}^+ p_{2} \right]\\ 
&I_{j \sigma}^h =\sum_{\sigma={\rm u, d}}\left(\gamma_{j,0}^+ -\gamma_{j,1}^-\right)p_\sigma -\gamma_{j,0}^- p_{0} + \gamma_{j,1}^+ p_{2},
\nonumber
\end{align}
where the transition rates for the heat current in \cref{currents} are given by:
\be
\gamma^{\pm}_{j,s}=\Gamma_{j} \int dE (E-eV_j) f^{\pm}(E-eV_j) P_j(E-\mu_{s}).
\ee
These rates take into account the heat transported in each particle transition. 

\subsection{Linear regime} 
\label{sec:linear}
By linearizing the electrical $I^{\rm e}_{i}$ and heat $I^{\rm h}_{i}$ currents at the $i$-th reservoir in response to the applied thermodynamical forces $\{V_j\, , \,T_j\}$ 
\begin{align}
\label{linearcurr}
I_i^{\rm e}&=\sum_j (G_{ij}V_j+L_{ij}T_j)\,,\\
I_i^{\rm h}&=\sum_j (M_{ij}V_j+K_{ij}T_j)\,,
\end{align}
we obtain the four conductance matrices that compose the Onsager matrix. Onsager-Casimir reciprocity relations dictate $G_{ij}=G_{ji}$,  $K_{ij}=K_{ji}$, and $L_{ij}=M_{ji}/T$. Additional relations imposed to the cross-conductances ($L_{ij}=L_{ji}$) arise in the case where transport occurs elastically. 

In our setup, the tunneling rates in~\cref{gammarates} describe inelastic processes. Hence, they introduce energy exchange with the environment. The energy of the two terminal system is hence not conserved. In the following we analyze the effect of inelasticicity on the thermal coefficients, namely $L_{ij}$ (or $M_{ji}$) and $K_{ij}$. 
%Due to this, we observe  both effects, namely (i) the cross-conductances are no longer symmetric $L_{\rm LR}\neq L_{\rm RL}$, and  $K_{\rm LL}\neq K_{\rm RR}$ whenever the environment is asymmetrically coupled to the quantum conductor. Such asymmetrical coupling can be performed either via a different junction capacitance (both $L$ and $K$ matrices are affected) for the two reservoirs or different tunneling rates at each barrier (only $K_{LL}\neq K_{RR}$). As a consequence the heat current becomes rectified in the presence of an asymmetrically coupled environment either by having $C_\rmL\neq C_\rmR$ or $\Gamma_\rmL\neq \Gamma_\rmR$

\subsection{Thermal coefficients}
Thermal rectification in an isoelectric ($V_\rmL=V_\rmR$) two terminal conductor occurs when the heat current becomes asymmetric on the reversal of the temperature gradient.
It has been discussed that is not possible in the linear regime for the heat current across the system.~\cite{Jiang2015} One has to take into account that energy is dissipated into the environment at the nanostructure. However in an experiment this quantity is not easy to detect. One would rather measure the heat current at each terminal. 

In this case, an apparent thermal rectification would be measured if
%The heat rectification phenomenon in a  two-terminal conductor in a isoelectric  ($V_\rmL=V_\rmR$) configuration occurs due to an asymmetry between the diagonal thermal conductance coefficients. Heat rectification, in general happens if heat currents for left and right contacts are not symmetric when the thermal gradient is inverted, i.e., more explicitly
\begin{equation}
\delta I_{\rmL\rmR}^{\rm h}=
I^{\rm h}_\rmL(V{=}0,\Delta T)-I^{\rm h}_\rmR(V{=}0,-\Delta T)\neq0.
\end{equation}
If we assume that the gradient is distributed between the two terminals, i.e. $\Delta T=\Delta T_\rmL-\Delta T_\rmR$, we get after linearizing the currents:
\begin{align}\label{heatrectification}
\delta I_{\rmL\rmR}^{\rm h}=
\left(K_{\rm LL}-K_{\rm RR}\right)\Delta T_\rmL+\left(K_{\rm LR}-K_{\rm RL}\right)\Delta T_\rmR.
\end{align}
The second term of the right-hand side of Eq.~(\ref{heatrectification}) vanishes due to the fulfilment of the Onsager relations. This however does not apply to the diagonal coefficients, $K_{\rm LL}$ and $K_{\rm RR}$. We will discuss below in which conditions these two coefficients become unequal in the presence of an environment, thus leading to asymmetric heat conduction.

\subsection{Thermoelectric coefficients}
It has been discussed that asymmetries of the $L_{ij}$ coefficients might improve the thermoelectric efficiency~\cite{Benenti2011}. This is the case for instance for broken time reversal symmetry in the presence of a magnetic field. Then, the efficiency at maximum power depends on the ratio $L_{ij}(B)/L_{ji}(-B)$.
%We additionally explore the thermoelectrical efficiency achieved at maximum power and how it depends on the environmental features. When the $L$ cross-conductance is no longer a symmetric matrix under a time-reversal inversion operation then, the thermoelectric efficiency is not uniquely defined by means of the $ZT$ figure of merit as currently \cite{Benenti2011}. Instead, such efficiency depends on the ratio $L_{ij}(B)/L_{ji}(-B)$ for instance when a magnetic field $B$ is present. Nevertheless, i
In our device $L$ becomes asymmetric under ``contact" inversion even in the absence of magnetic field.  

In this case, our system acts as an engine which generates a finite power when the thermally activated current flows against a voltage gradient. Important coefficients of performance are the maximum generated power and the efficiency at maximum power. Let us especifiy a configuration where $T_\rmL>T_\rmR$, and an applied voltage $V_\rmR-V_\rmL=\Delta V$. The extracted power 
\begin{equation}
P= -I_\rmL(\Delta V) \Delta V = I_\rmR(\Delta V)\Delta V
\end{equation}
is maximized for some drop voltage $\Delta V=V_{\rm m}$, giving: 
\begin{equation}
%P_{\rm max} =  -G_{\rm LR} V^{2}_{\rm m}-(L_{\rm LR} \Delta T_\rmR+ L_{\rm LL} \Delta T_\rmL) V_{\rm m}
V_{\rm m}=-(L_{\rm LR} \Delta T_R + L_{\rm LL} \Delta T_\rmL)/(2G_{\rm LR}),
\end{equation}
%with $V_{\rm m}=-(L_{\rm LR} \Delta T_R + L_{\rm LL} \Delta T_\rmL)/(2G_{\rm LR})$ 
which results in a maximum power:
\begin{equation}\label{pmax}
P_{\rm max}=\frac{(L_{\rm LR} \Delta T_\rmR + L_{\rm LL} \Delta T_\rmL)^2}{4G_{\rm LR}}.
\end{equation}
Finally the efficiency at maximum power $\eta_{\rm max}$ is easily computable from 
%\begin{widetext}
%\begin{equation}
%\eta_{max} = \frac{P_{max}}{-I^h_{L}(V_m)} =\frac{(L_{LR} \Delta T_R + L_{LL} \Delta T_L)^2}{2 M_{LR}(L_{LR} \Delta T_R + L_{LL} \Delta T_L)- 4G_{LR}(K_{LR} \Delta T_R + K_{LL} \Delta T_L)}
%\end{equation}
%\end{widetext}
\begin{equation}\label{effmax}
\eta_{\rm max} = \frac{-P_{\rm max}}{I^h_{\rmL}(V_{\rm m})+I^h_{\rm E}(V_{\rm m})},
\end{equation}
where one has to take into account the heat current injected by the environment, $I^{\rm h}_{\rm E}$.% which can promote or block energy transport.
%This efficiency takes a much simpler form if we allow symmetrical thermal drops as $\Delta T_L=-\Delta T_R =\Delta T/2$, then
%$\eta_{max} = (\delta T/4) (L_{LR}-L_{RR})^2/(M_{LR}(L_{LL}-L_{LR})-2G_{LR}(K_{LL})-K_{LR})$.% where $\eta_C =\Delta T/T_0$ being $T_0$ the background temperature for both reservoirs (that could be different from the environmental temperature). 

In order to carry a more meaningful study of the setup efficiency, we also study the Carnot efficiency for this setup. In our case we need to take into account that heat is being injected by the environment and thus the Carnot efficiency is not simply $\eta_C=1-T_\rmR/T_\rmL$. We define the Carnot efficiency at the reversible point where $\frac{I^{\rm h}_{\rmL}}{T_\rmL}+\frac{I^{\rm h}_{\rmR}}{T_\rmR}+\frac{I^{\rm h}_{\rm E}}{T_0}=0$ (zero entropy production), which results in:
\begin{equation}
\eta_C=1+\left(\frac{I^h_{\rmL}}{I^h_{\rmR}} \frac{1-T_0}{T_R}-\frac{T_0}{T_L}\right)^{-1}.
\end{equation}
This efficiency gives a measure of the performance of our setup. 

%********************************
\section{Single occupancy}
\label{sec:spinless}
%********************************
We can make some analytical progress by considering a simplified situation. Let us assume the limit $E_C\gg k_{\rm B}T_0$, such that the quantum dot can be occupied by a single electron at a time. %In the absence of a magnetic field, the spin of the electron can then be ignored. 
It will later help to understand the numerical results for the general configuration presented in Sec.~\ref{sec:spinful}. 
 In this case, the charge current simply reads:
\be
\label{currSpinless}
I_\rmL^\rme=e\frac{\Gamma_{\rmL,0}^+\Gamma_{\rmR,0}^--\Gamma_{\rmL,0}^-\Gamma_{\rmR,0}^+}{\Gamma_{\rmL,0}^++\Gamma_{\rmR,0}^-+\Gamma_{\rmL,0}^-+\Gamma_{\rmR,0}^+},
\ee
with $I_\rmR^\rme=-I_\rmL^\rme=I$.  We consider the isoelectric case (we define $\Delta\mu^0=\epsilon_d+e^2/2C$) and compute the linearized charge current in contact $l$ by linearizing tunneling rates  as follows
%\begin{align}
%\label{fermiexpand}
%f(E,kT_l){=}f(E,kT){+}\frac{E\Delta T_l}{kT^2}f(E,kT)(1-f(E,kT)).
%\end{align}
%We get:
\begin{align}
\Gamma_{l,0}^+&= \Gamma_{l}\left(g_{l}^{(0)}+\frac{\Delta T_l}{T}g'_{l}\right)\\
\Gamma_{l,0}^-&=e^{\Delta\mu^0/kT}\left(\Gamma_l^+-\frac{\Delta T_l}{T} \Gamma_{l}{g}_{l}^{(1)}\right),
\end{align}
where we have introduced the following integrals:
\begin{align}
g_{l}^{(n)}&=\int dE\left(\frac{E+\Delta\mu^0}{kT}\right)^nf^+(E+\Delta\mu^0)P_l(E),\nonumber\\
g_{l}^{'n}&=\frac{1}{kT}\int dE\left(\frac{E{+}\Delta\mu^0}{kT}\right)^nf^+(E{+}\Delta\mu^0)\nonumber\\
&\times f^-(E{+}\Delta\mu^0)P_l(E),\nonumber
\end{align}
which only depend on the corresponding terminal through the $P_l(E)$ function, $l=\rmL,\rmR$. In the following, we write $f^\pm(E)$ for $f^\pm(E,T)$.
%For the expansion of $\Gamma_l^-$, we have used:
%\begin{align}
%1-f(E+\Delta \mu^0)=
%e^{(E+\Delta \mu^0)/kT_l}f(E+\Delta \mu^0).
%\end{align}
%We then expand the Fermi function in the r.h.s. as in Eq.~\eqref{fermiexpand}, and the exponential: $e^{(E+\Delta \mu^0)/kT_l}\approx\left(1-\frac{E+\Delta \mu^0}{kT^2}\Delta T_l\right)e^{(E+\Delta \mu^0)/kT}$. Finally, one replaces $P(-E)=e^{-E/kT}P(E)$. 

Replacing them into Eq.~\eqref{currSpinless} we obtain the Seebeck coefficients:
\begin{align*}
L_{\rm LR}&=-\frac{e \Gamma_{\rmL} \Gamma_{\rmR}}{\sum_l \Gamma_{l}g_l^{(0)}}e^{\Delta\mu^0/kT}f^+(\Delta\mu^0)g_\rmL^{(0)}g_\rmR^{(1)}\\
L_{\rm RL}&=-\frac{e \Gamma_{\rmL} \Gamma_{\rmR}}{\sum_l \Gamma_{l}g_l^{(0)}}e^{\Delta\mu^0/kT}f^+(\Delta\mu^0)g_\rmR^{(0)}g_\rmL^{(1)}.
\end{align*}
The asymmetry in the thermoelectric coefficients is hence:
\be \label{difL}
L_{\rm LR}-L_{\rm RL}\propto g_\rmL^{(0)}g_\rmR^{(1)}-g_\rmR^{(0)}g_\rmL^{(1)}=X_{\rm RL}^{(1)},
\ee
with
\begin{align}\label{chill}
X_{ll'}^{(n)}={\int}&{dEdE'}\left(\frac{E{+}\Delta\mu^0}{kT}\right)^nf^+(E{+}\Delta\mu^0)f^+(E'{+}\Delta\mu^0)\nonumber\\
&\times[P_l(E)P_{l'}(E')-P_l(E')P_{l'}(E)].
\end{align}
We can then explicitly relate the asymmetry of the Seebeck coefficients \eqref{difL} to the inhomogeneous influence of the environment on the tunneling processes through each barrier, $P_l(E)$. This occurs when $C_\rmL\neq C_\rmR$: both the mean and variance of the distributions become different, cf.~\cref{PjofE}. Hereafter, we parametrize the asymmetry of the tunnel barrier capacitances with
\begin{equation}
\kappa=\frac{C_\rmL-C_\rmR}{C}.
\end{equation} 
Then, for a finite $\kappa$ it follows that  $P_\rmL(E)\neq P_\rmR(E)$. Note that  $X_{ll'}^{(n)}$ is independent of the tunneling rates $\Gamma_\rmL$, and $\Gamma_\rmR$.

The functions $X_{ll'}^{(n)}$ (and therefore the asymmetry) depend on the ovelapping of the distributions $P_\rmL(E)$ and $P_\rmR(E)$, which depends on $C$. For very small capacitances, $C\rightarrow0$, the two distributions are narrow and they do not overlap. In the opposite limit, they are so wide that their difference is tiny. In between we can therefore find optimal values of $C$ for which the effect of the environment is maximal.

\begin{figure}[t!]
{\centering
\includegraphics[width= \linewidth,clip]{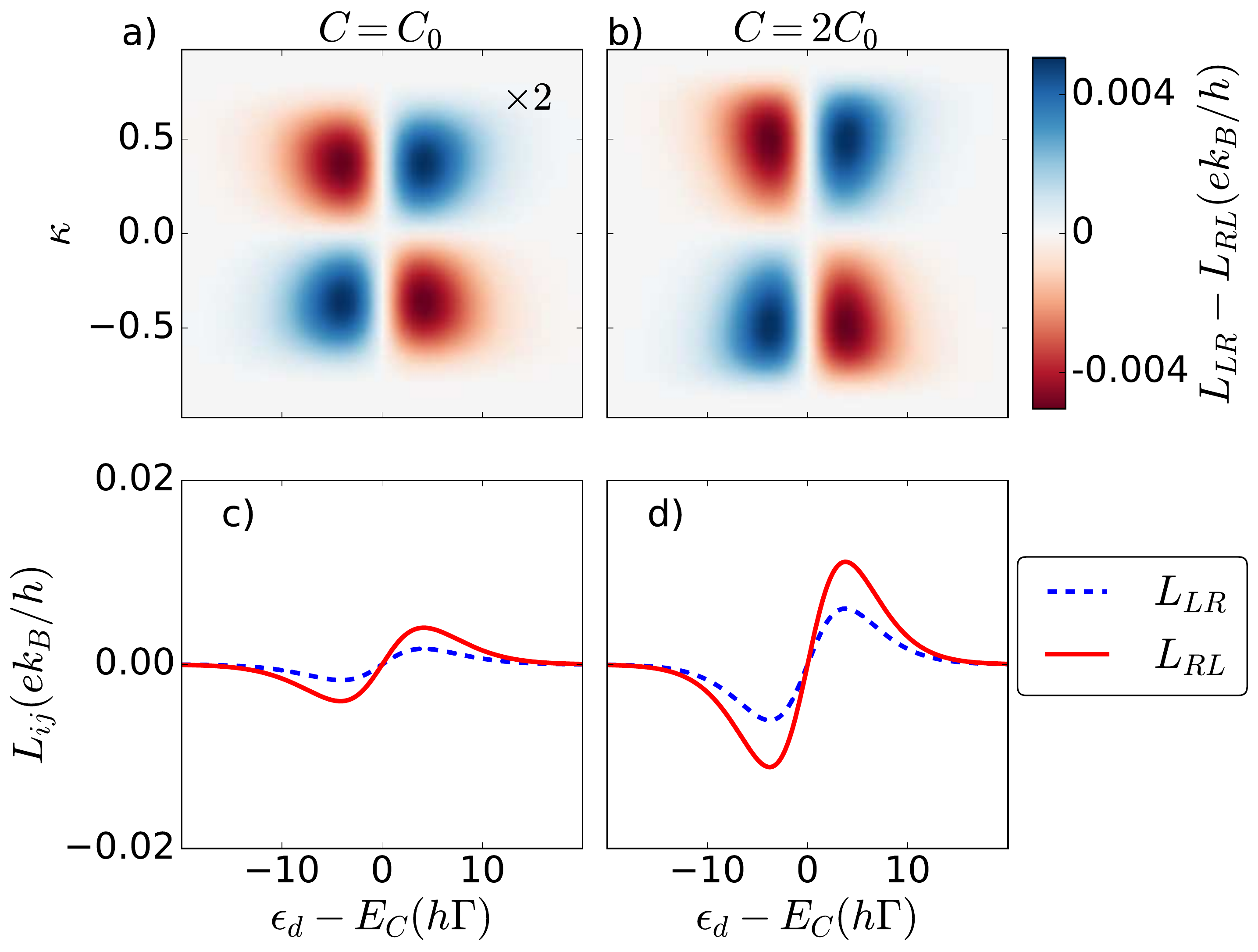}
}
\caption{\small Electrothermal asymmetry $L_{\rm LR}- L_{\rm RL}$ versus dot energy level position $\epsilon_d-E_C$ 
and capacitance asymmetry $\kappa$ for different values of the total capacitance (a) $C=C_0$, and (b) $C=2C_0$. (c) 
and (d) show cuts of $L_{\rm LR}$, and $L_{\rm RL}$ when $\kappa=0.5$, for the values of $C$ considered above. We have taken $k_B T_0=2 h\Gamma_0$ and $\Gamma_L=\Gamma_R=\Gamma_0/2$. The crossed terms $L_{\rm LR}$, and $L_{\rm RL}$ differ between themselves for $\kappa \neq 0$, i.e. ($C_\rmL\neq C_\rmR$).}
\label{fig:Lasym}
\end{figure}

These observations are reflected in  \cref{fig:Lasym}. There we have represented $L_{\rm LR}-L_{\rm RL}$ for two different values of the total capacitance $C=C_0$ [\cref{fig:Lasym}(a)] and $2 C_0$ [\cref{fig:Lasym}(b)], with $C_0=e^2/10 h\Gamma_0$, as a function of the energy level of the dot $\epsilon_{\rm d}$ and $\kappa$. We vary the symmetry of the capacitances described by  $\kappa$ but not the total capacitance $C$ which is kept constant. As seen in \cref{fig:Lasym} the difference  $L_{\rm LR}-L_{\rm RL}$  is zero for $\kappa=0$. This means that an environment coupled symmetrically to both barriers is not able to break this symmetry and therefore, inelastic scattering is not a sufficient condition in order to break this symmetry in the Onsager matrix. However, we observe that for a finite $\kappa$ both contributions, $L_{\rm LR}$, $L_{\rm RL}$ differ, see \cref{fig:Lasym}(c) and (d). Both have the expected saw-tooth lineshape~\cite{beenakker1992}. Remarkably, $L_{\rm LR}-L_{\rm RL}$ reverses its sign when  $\epsilon_{\rm d}-E_C$ and $\kappa$ are tuned are shown in the upper pannel of \cref{fig:Lasym}. It changes sign when the asymmetry is inverted (i.e. when $\kappa$ changes sign) and at the particle-hole symmetry point $\epsilon_{\rm d}-E_C=0$, where also $L_{ij}=0$. 

The maximal value of  $L_{\rm LR}-L_{\rm RL}$ (at a finite value of $\kappa$) depend strongly on the total capacitance $C$. The difference $L_{\rm LR}-L_{\rm RL}$ increases with $C$. A plausible argument for such behavior is obtained by looking at the integrand of \cref{chill} that depends on the factor $P_\rmR(E)P_{\rmL}(E')-P_\rmR(E')P_{\rmL}(E)$ and accounts for the overlap between two Gaussian functions each centered at energy positions that depend on $\kappa_\rmL^2 E_C$ and $\kappa_\rmR^2 E_C$ (even though their widths depends on the same factor, as well), by making $C$ high yields an overlapping between the two Gaussian functions and enhances the value for $X_{\rm RL}^{(1)}$. For such reason, $L_{\rm LR}-L_{\rm RL}$ reaches lower values whenever $C$ is smaller. 

A similar analysis is performed for the thermal coefficients.
% Since we are interested in the thermal coefficients, $K_{ll'}$, we consider the zero-voltage condition $\Delta\mu^0=\varepsilon-\mu$,  with the position of the level $\varepsilon=\varepsilon_0+\kappa_L\kappa_RE_C$ and the Fermi energy $\mu$. 
Expanding the heat rates we obtain:
\begin{align}
H_l^+&= \Gamma_l\left(kTg_l^{(1)}+g''_lk\Delta T_l\right)\nonumber\\
H_l^-&=e^{\Delta\mu^0/kT}\left(H_l^+- \Gamma_lg^{(2)}_lk\Delta T_l\right).\nonumber
\end{align}
With these relations, we get the expression for the (diagonal) linear thermal asymmetry, $\delta K=K_{\rm LL}-K_{\rm RR}$. It can be separated in two contributions, $\delta K=\delta K_\Gamma+\delta K_\kappa$, with:
\begin{align}\label{asymK}
\delta K_\Gamma&=e^{\Delta\mu^0/kT}\frac{f(\Delta\mu^0)}{\sum_l \Gamma_{l}g_l^{(0)}}\left( \Gamma_\rmL^2\Theta_\rmL- \Gamma_\rmR^2\Theta_\rmR\right)\\
\delta K_\kappa&=e^{\Delta\mu^0/kT}\frac{f(\Delta\mu^0)}{\sum_l \Gamma_{l}g_l^{(0)}}\Gamma_\rmL \Gamma_\rmR X_{\rm LR}^{(2)}.
\end{align}
Here we have defined:
\begin{align}
\Theta_{l}{=}&-k_{\rm B}T\left[\left(g_l^{(1)}\right)^2-g_l^{(2)}g_l^{(0)}\right],
%\nonumber\\
%=&{\int}{dEdE'}(E{-}E')(E{+}\Delta\mu^0)f^+(E{+}\Delta\mu^0)\nonumber\\
%&\times f^+(E'{+}\Delta\mu^0)P_l(E)P_{l}(E'),
\end{align}
which depends on the tunneling events through a single barrier. We can hence distinguish two sources of recification: $\delta K_\Gamma$ becomes non zero when asymmetric tunneling barriers are considered, i.e. $\Gamma_\rmL\neq \Gamma_\rmR$ or when $\kappa\neq 0$. It relates to the different time scales that an electron stays in contact with the environment when tunneling from the left or from the right reservoir, as we discuss below. However, $\delta K_\kappa$ is is intrinsically dependent on the dynamic coupling to the environment: it is only non zero when the capacitances for each tunneling barrier are different. 

\begin{figure}[t!]
{\centering
\includegraphics[width= \linewidth,clip]{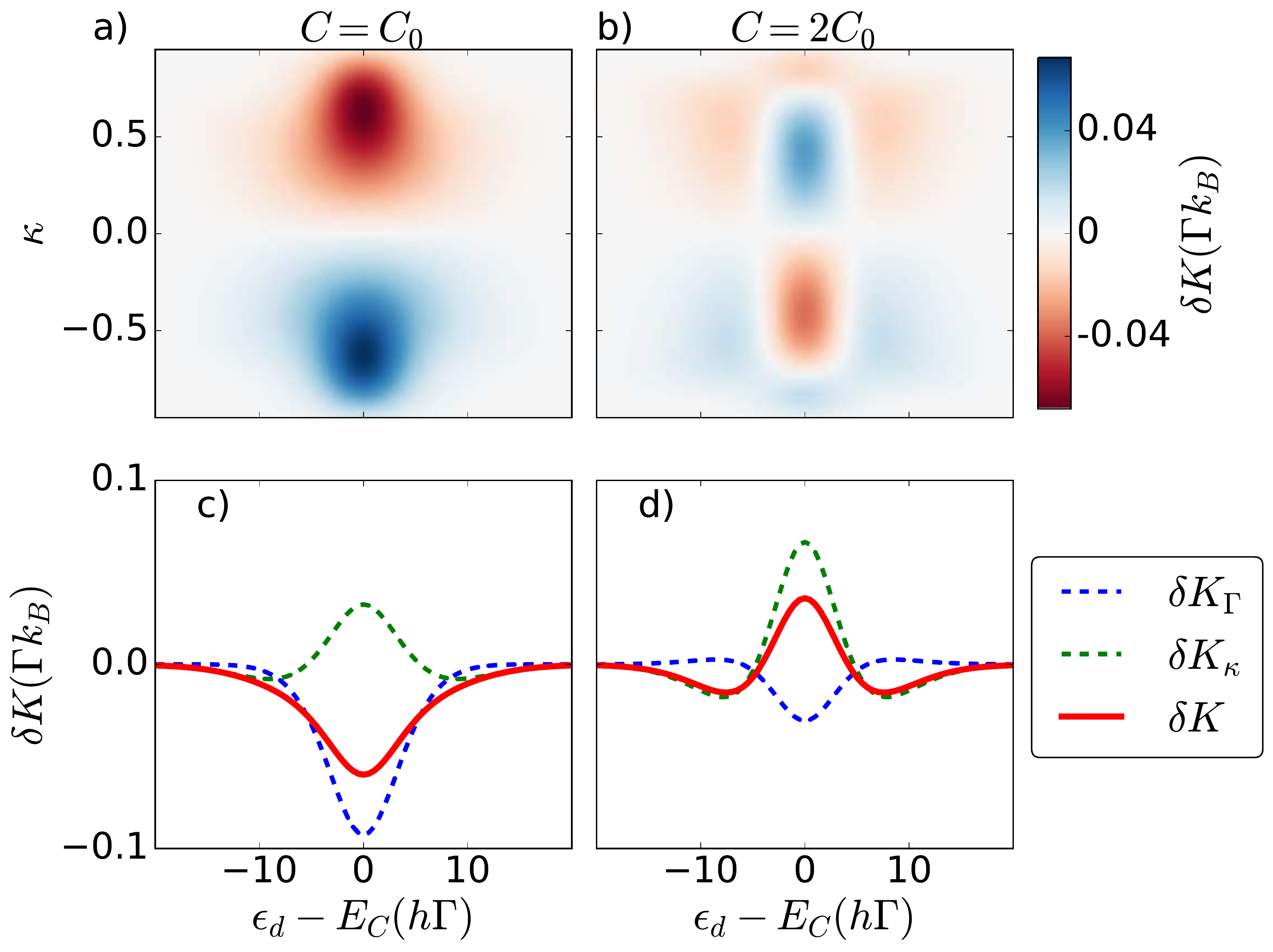}
}
\caption{\small a) b) Thermal asymmetry $K_{\rm LL}- K_{\rm RR}$ versus dot energy level position $\epsilon_d$ and capacitance asymmetry $\kappa$ for different values of the total capacitance $C=C_0,2 C_0$. Heat rectification $K_{\rm LL}- K_{\rm RR} \neq 0$ for $\kappa\neq 0$, i.e. ($C_\rmL\neq C_\rmR$). c) d) Total $\delta K$ (red line), tunnel $\delta K_\Gamma$ (dashed blue line) and capacitance $\delta K_\kappa$ (dashed green line) thermal asymmetries evaluated at $\kappa=0.5$ for two different capacitances $C=C_0=2 C_0$. The change of sign of $\delta K_\Gamma$ which implies the change of sign of the total $\delta K$ is clearly observed. Parameters are the same as in \cref{fig:Lasym}.  }
\label{fig:deltak2}
\end{figure}

The role of a capacitance asymmetry on the heat rectification $\delta K$ is plotted in \cref{fig:deltak2}.  There, we show $K_{\rm LL} - K_{\rm RR}$ versus the dot gate potential $\veps_{\rm d}-E_C$ when the asymmetry in the capacitances $\kappa$ is tuned. We observe that heat rectification reverses its sign when $\kappa$ does too. Importantly, we find different behaviours depending on the total capacitance $C$. For large enough $C$, $\delta K$ changes sign with the position of the level, cf. Fig.~\ref{fig:deltak2}(b). This is due to a change in the relative contribution of the two terms, $\delta K_\Gamma$ and $\delta K_\kappa$, as shown in the lower panels in Fig.~\ref{fig:deltak2}. This effect introduces an additional way of controlling the heat flows through the device, depending on the position of the different mean values of the $P_j(E)$ distributions with respect to the Fermi energy.

For small values of $C$, the $P(E)$ functions of the two contacts differ greatly and therefore $\delta K_\Gamma$ is big (and negative for $\kappa>0$), see \cref{fig:deltak2}(c). Therefore it dominates over $\delta K_{\kappa}$ for $C=C_0$. Instead, for high values of $C$ we showed that $X_{\rm LR}^{(n)}$ enhances strongly (the two environmental $P(E)$-functions overlap when $C$ is high, leading to a greater contribution) and on top of that, $\delta K_{\Gamma}$ has now decreased, see \cref{fig:deltak2}(d). Therefore, from the competition between the two terms and its dependence with $\kappa$ one obtains a nontrivial dependence of the heat rectification with $C$.

\begin{figure}[t!]
{\centering
\includegraphics[width= \linewidth,clip]{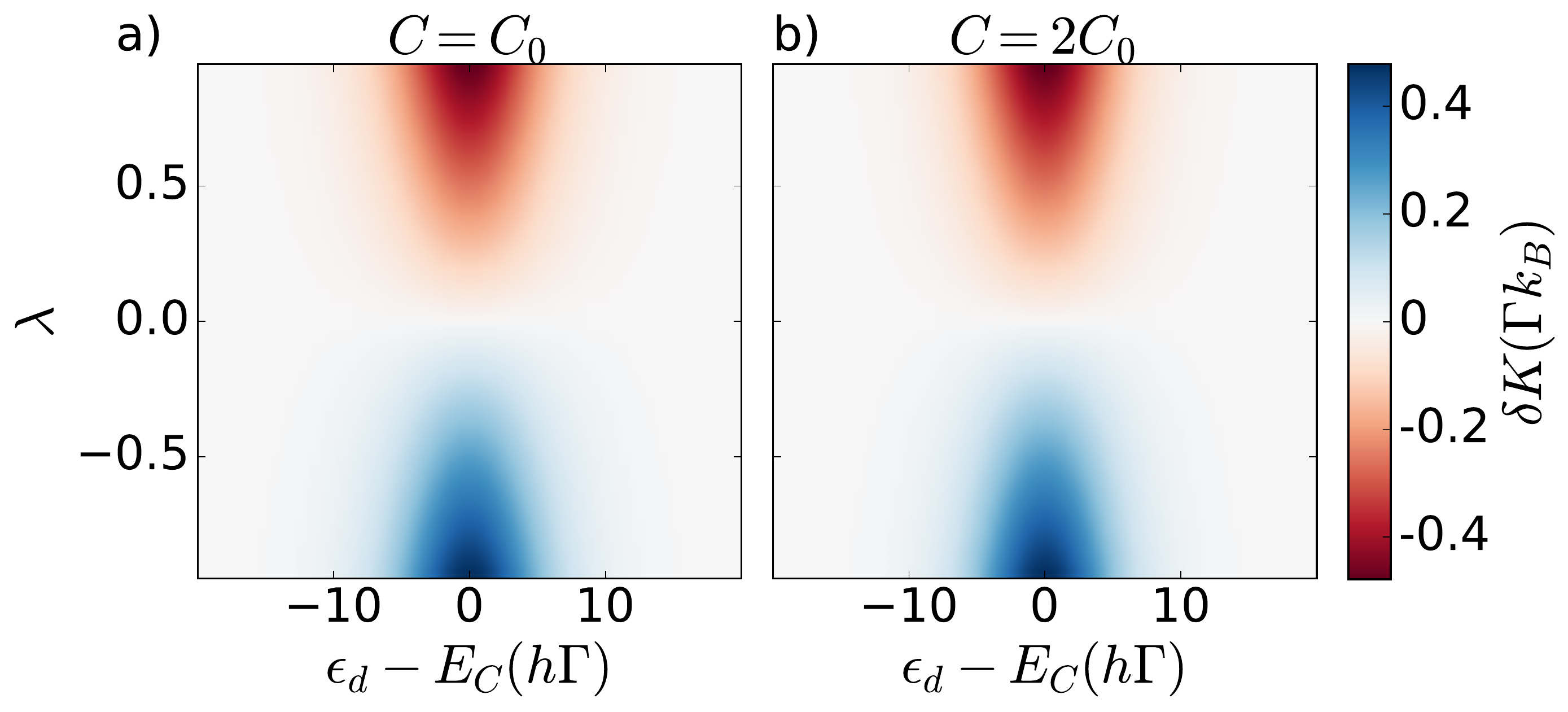}
}
\caption{\small Thermal asymmetry $K_{\rm LL}- K_{\rm RR}$ versus dot energy level position $\epsilon_d$ and barrier asymmetry $\lambda$ for different values of the total capacitance $C=C_0,2 C_0$. Heat rectification $K_{\rm LL}- K_{\rm RR} \neq 0$ for $\gamma\neq 0$, i.e. ($\Gamma_\rmL\neq \Gamma_{\rmR}$). Parameters are the same as in \cref{fig:Lasym}.}
\label{fig:KasymG}
\end{figure}

Now we explore the heat rectification arising from different tunneling amplitudes, see \cref{fig:KasymG}. There, we show $K_{\rm LL} - K_{\rm RR}$ versus the dot gate potential $\veps_{\rm d}-E_C$ when the asymmetry in the tunneling rate, parametrized as
\begin{equation}
\lambda=\frac{\Gamma_\rmL-\Gamma_\rmR}{\Gamma}
\end{equation}
is tuned, with $\Gamma=\Gamma_\rmL+\Gamma_\rmR$.
Heat rectification stems from purely different kinetic couplings. In this case $\delta K_\Gamma$ is the only source of rectification in the heat flow. In that case, the same $P(E)$ function represent the same environment for the two barriers. However, by allowing $\Gamma_\rmL\neq \Gamma_{\rmR}$ electrons traversing the left barrier spend shorter times in contact with the environment (and therefore the time where energy-exchange processes are taken is shorter) than electrons at the other barrier. Therefore environmental assisted tunneling transitions are effectively different for both junctions even though they are performed with the same environment, with $P_\rmL(E)=P_\rmR(E)$. Results for $K_{\rm LL}{-}K_{\rm RR}$ at $C_\rmL=C_\rmR=C/2$ for different values of the barrier asymmetry $\lambda$  are shown in \cref{fig:KasymG}. In this figure we observe that indeed for $\lambda \neq 0$ there is an asymmetry of the thermal coefficients which changes sign with $\lambda$. It is also observed that the shape of the asymmetry does not depend strongly on the total capacitance $C$, as expected, since the source for such heat rectification depends essentially on $\lambda$ and it does not have an electrostatic origin. Even so, the total value of the rectification decreases with the capacitance. 

Finally, we focus on how the thermoelectric performance of the device is affected by the environment. Discrete levels in quantum dot systems are known to give a high performance for being ideal energy filters. In terms of efficiency at maximum power, they reach the Curzon-Ahlborn efficiency~\cite{Esposito2009} which in the linear regime is $\eta_{\rm CA}=\eta_C/2$~\cite{Curzon1975}.
%{Humphrey2005a}. 
As discussed above, energy filtering is harmed by the occurrence of inelastic scattering. Hence, the efficiency is expected to be smaller~\cite{Nakpathomkun2010}. 

\begin{figure}[t!]
{\centering
\includegraphics[width= \linewidth,clip]{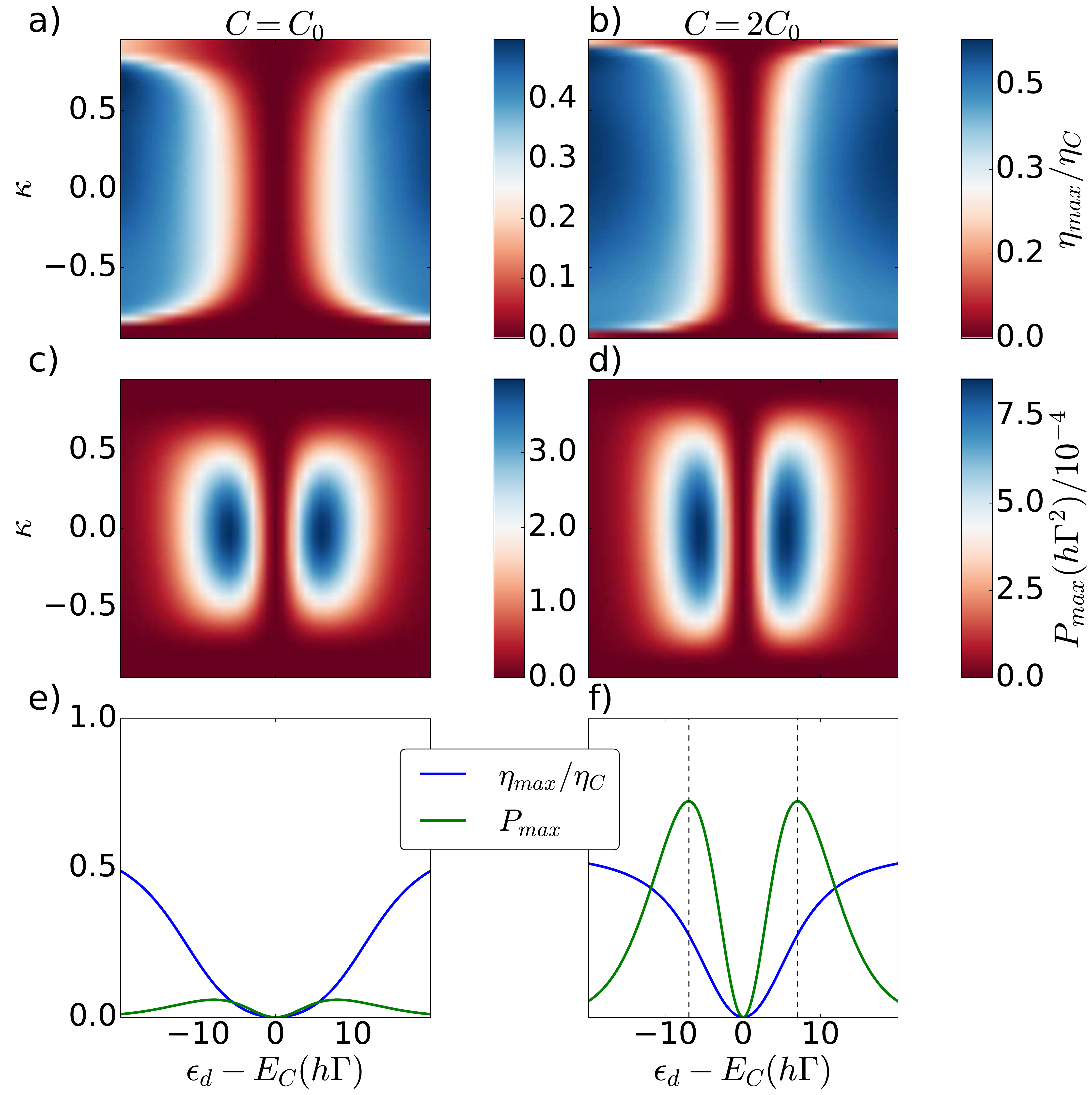}
}
\caption{\small Efficiency at maximum power $\eta_{\rm max}$ in units of the Carnot efficiency $\eta_{\rm C}$ computed at maximum power versus the dot level position $\epsilon_{\rm d}$ and the asymmetry parameter $\kappa$, for different capacitances: (a) $C=C_0$ and (b) $C=2C_0$. (c), (d): Maximum power $P_{\rm max}$ extracted corresponding at the efficiency $\eta$ showed in (a) and (b), respectively. Cuts of the previous curves for $\kappa=1/2$ are presented in (e) and (f), for clarity. Parameters are the same as in \cref{fig:Lasym}.} 
\label{fig:eff}
\end{figure}

This is indeed what we observe in \cref{fig:eff}, which shows the efficiency at maximum power $\eta_{\rm max}$ [\cref{effmax}] and the maximum power [\cref{pmax}] as functions of the dot gate position $\epsilon_{\rm d}$ and the asymmetry parameter $\kappa$. We find there that the system reaches efficiencies close to the $\eta_{\rm CA}$ bound, or even larger, cf. \cref{fig:eff}(f).
This happens for large values of the dot level position and at $\kappa>0$. The efficiency is increased whenever $C_\rmL> C_\rmR$ since this coupling favors the injection of heat from the environment to the right lead effectively helping electrons overcome the bias potential, therefore less heat is needed from the left reservoir to extract the same power and hence the increase in efficiency. Unfortunately, at these configurations the output power is strongly suppressed, as displayed in \cref{fig:eff}. Nevertheless, the highest $P_{\rm max}$ can be extracted at reasonably high efficiencies $\eta_{max} \sim \eta_C/3$, see  \cref{fig:eff}(f). Therefore, we observed environmental enhanced efficiencies (as compared with a perfect energy filter~\cite{Esposito2009}, which are bound by the Curzon-Ahlborn limit).% which unfortunatelly correspond to situations with reduced power extraction. 
%Therefore, we see that although an efficiency higher can be achieved in the presence of  an environment, the maximum power that is extracted is lower that for the case without environment. \cite{Esposito2009}. 
%Besides, in the case without environment the highest maximum power reached is always attained for higher efficiencies than in the case with it. 

We also note that the efficiency is strongly dependent on the details of the coupling to the environment, evidenced by the comparison of Figs.~\ref{fig:eff}(e) and (f). Larger $C$ not only gives larger power, it also generates it at much larger efficiencies, as compared with lower $C$. Hence, we expect that the engineering of the environmental fluctuations (by considering e.g. non-ohmic impedances) could result in devices with enhanced thermoelectric performances.

\section{Double occupancy}
\label{sec:spinful}
In the light of the results presented in the previous section for the reduced Hilbert space with up to one electron, we come back to the general configuration allowing for double occupancy. That is, now we consider $|0\rangle $,  $|{\rm u}\rangle$, $|{\rm d}\rangle$ and $|2\rangle$. Analytical results for the double occupancy case are cumbersome so we restrict ourselves to  present our numerical simulations. The results presented in this section show the two differences:  $L_{\rm LR}-L_{\rm RL}$ and $\delta K$. 

\begin{figure}[t!]
{\centering
\includegraphics[width= \linewidth,clip]{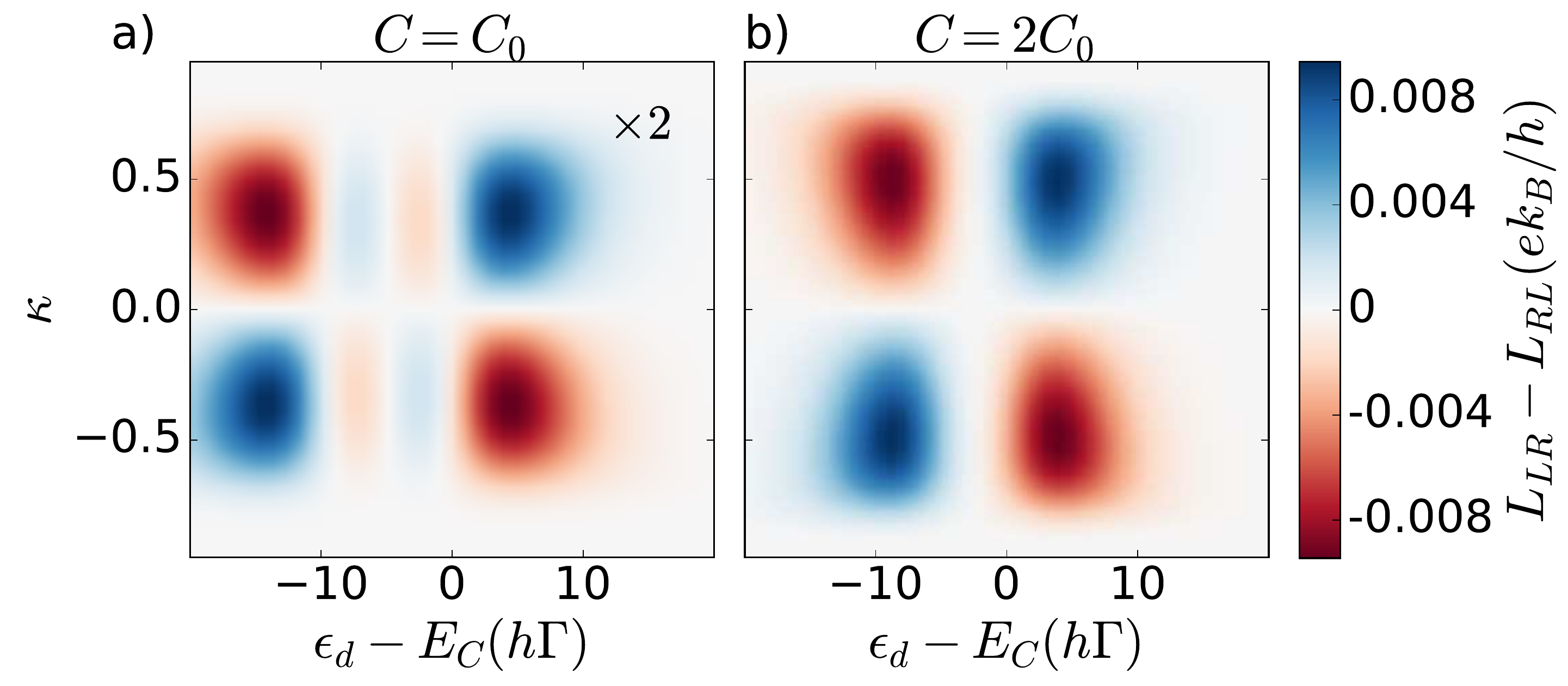}
}
\caption{\small Double occupancy electrothermal asymmetry $L_{\rm LR}- L_{\rm RL}$ versus dot energy level position $\epsilon_{\rm d}-E_C$ and capacitance asymmetry $\kappa$ for different values of the total capacitance $C=C_0,2 C_0$. The crossed terms $L_{LR}$, and $L_{\rm RL}$ differ between themselves for $\kappa\neq 0$, i.e. ($C_\rmL\neq C_\rmR$). Parameters are the same as in \cref{fig:Lasym}.}
\label{fig:DOLasym}
\end{figure}

Firstly, we present the results for $L_{\rm LR}-L_{\rm RL}$ in \cref{fig:DOLasym}. As expected,  such difference arises only with an electrostatic asymmetry coupling, i.e.,  when $\kappa \neq 0$. Now, a second peak for $L_{\rm LR}-L_{\rm RL}$ appears corresponding to the charging of the system with a second electron. The two saw-tooth oscillations are separated by the charging energy $\mu_1-\mu_0=10 h\Gamma$, see \cref{fig:DOLasym}(a). Notice that when the charging energy becomes sufficiently small, i.e. for $C=2C_0$, the two features come closer and the inner oscillations are no longer visible. The behaviour then resembles the one obtained in the opposite limit (large charging energy) in \cref{sec:spinless}. %the results for the single occupancy case (apart from a factor 2 coming from an almost spin degeneracy when the charging energy is negligible). 

\begin{figure}[t!]
{\centering
\includegraphics[width= \linewidth,clip]{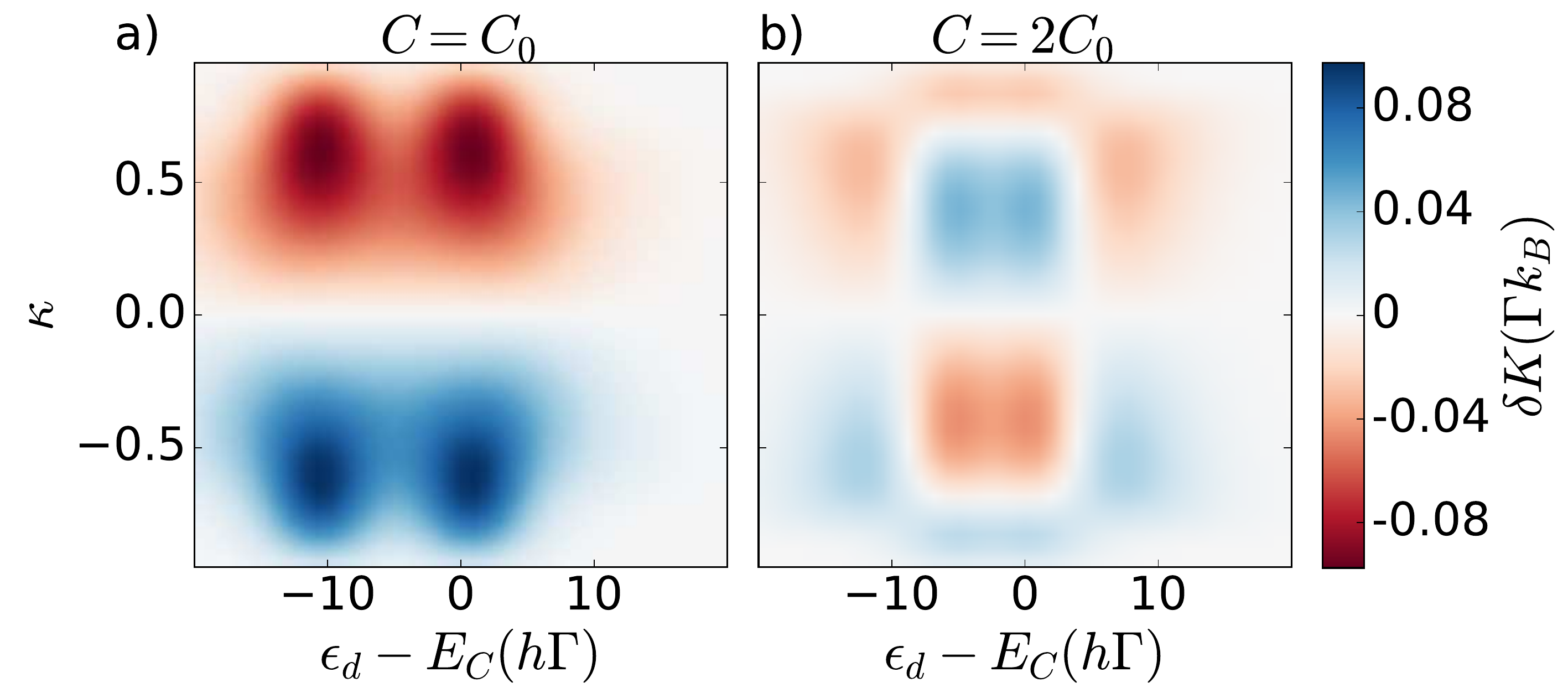}
}
\caption{\small Double occupancy thermal asymmetry $K_{\rm LL}- K_{\rm RR}$ versus dot energy level position $\epsilon_{\rm  d}$ and capacitance asymmetry $\kappa$ for different values of the total capacitance $C=C_0,\ 2 C_0$. Heat rectification $K_{\rm LL}- K_{\rm RR} \neq 0$ for $
\kappa\neq 0$, i.e. ($C_\rmL\neq C_\rmR$). Same parameters as in \cref{fig:Lasym}.}
\label{fig:DOKasym}
\end{figure}

The heat rectification for the double occupancy case is plotted in Fig. \ref{fig:DOKasym}. We observe the double peak structure for a low capacitance $C=C_0$ due to the two different energy levels available. As $C$ increases there are two main effects (i) the overall double peak tends to disappear since the charging energy diminishes, and (ii) sign changes appear in the heat rectification (for fixed $\kappa$) due a greater weight of $\delta K_\kappa$ against $\delta K_\Gamma$ when $C$ increases. Therefore the sign of the heat rectification is mainly given by $\delta K_\kappa$ for a large capacitance $C$ whereas $\delta K_\Gamma$ determines the heat rectification sign when $C$ is small. 
%By comparing with the results shown in \cref{sec:spinless} for large charging energies, we observe that the magnitude of the rectification coincides for large and small charging energies (cf. Figs.~\ref{fig:deltak2} and \ref{fig:DOKasym}(b), respectively), but is enhanced for intermediate capacitances, cf. \cref{fig:DOKasym}(a).
%Again, for the four state case ($C=2C_0$) a factor two increase in the rectification with respect to that of the single occupancy case is observed. This is not the case when the capacitance is smaller due to the fact that the different available energy levels are far apart in energy, meaning that electrons transported through the dot do not all carry a similar energy.

\section{Conclusions}
\label{sec:conclusions}
In closing, we have analyzed the effect of inelasticity introduced by an electromagnetic environment on transport through a conductor (a quantum dot). We have found that even in the absence of a magnetic field, an asymmetric energy exchange with the environment can break symmetries of the transport coefficients which has a consequence the apparent occurrance of heat rectification in the linear regime. Furthermore, we have shown that heat injected from the environment can either improve or diminish the efficiency at maximum power output of the device when used as an engine. Efficiencies close to the Curzon-Ahlborn are attainable even if at vanishing power output. 

We have considered here the case of a high impedance environment. Other kind of interactions will have different impact on the performance of the system. The experimental hability to engineer the electromagnetic environment~\cite{Altimiras2014} opens the way to improve the control of thermal flows in mesoscopic conductors.

\begin{acknowledgments}
We acknowledge discussions with  D. S\'anchez. G.R. and R.L were supported by MINECO Grants No. FIS2014-52564. R.S. acknowledges financial support from the Spanish Ministerio de Econom\'ia y Competitividad via grant No. MAT2014-58241-P. We also thank the COST Action MP1209 ``Thermodynamics in the quantum regime".
\end{acknowledgments}

\bibliographystyle{apsrev4-1}
\bibliography{Articleinelastic2}

\end{document}